# Laser driven miniature diamond implant for wireless retinal prostheses

Arman Ahnood[1,2], Ross Cheriton[3,4], Anne Bruneau[4], James A. Belcourt[1], Jean Pierre Ndabakuranye[1], William Lemaire[5], Rob Hilkes[4], Réjean Fontaine[5], John P.D. Cook[6], Karin Hinzer[6], Steven Prawer[1]

[1] *School of Physics, University of Melbourne, Parkville, Australia.*

[2] *School of Engineering, RMIT University, Melbourne, Australia.*

[3] *National Research Council of Canada, Ottawa, Canada.*

[4] *iBIONICS, Ottawa, Canada.*

[5] *Interdisciplinary Institute for Technological Innovation, Université de Sherbrooke, Sherbrooke, Canada.*

[6] *Centre for Research in Photonics, University of Ottawa, Ottawa, Canada.*

The design and benchtop operation of a wireless miniature epiretinal stimulator implant is reported. The implant is optically powered and controlled using safe illumination at near-infrared wavelengths. An application-specific integrated circuit (ASIC) hosting a digital control unit is used to control the implant's electrodes. The ASIC is powered using an advanced photovoltaic (PV) cell and programmed using a single photodiode. Diamond packaging technology is utilised to achieve high-density integration of the implant optoelectronic circuitry, as well as individual connections between a stimulator chip and 256 electrodes, within a 4.6 mm × 3.7 mm × 0.9 mm implant package. An ultra-high efficiency PV cell with a monochromatic power conversion efficiency of 55% is used to power the implant. On-board photodetection circuity with a bandwidth of 3.7 MHz is used for forward data telemetry of stimulation parameters. In comparison to implants which utilise inductively coupled coils, laser power delivery enables a high degree of miniaturisation and lower surgical complexity. The device presented combines the benefits of implant miniaturisation and a flexible stimulation strategy provided by a dedicated stimulator chip. This development provides a route to fully wireless miniaturised minimally invasive implants with sophisticated functionalities.

# 1. Introduction

Advances in the complementary metal-oxide semiconductor (CMOS) technology have broadened the scope and capabilities of bioelectronic implants [1–5]. Miniaturised application specific integrated circuits (ASIC) with significant capabilities in terms of signal processing, speed, and computational power are now within the realm of possibilities. These technological developments have coincided with advances in minimally invasive surgical interventions such as keyhole surgical procedures. The combination of these breakthroughs has supported a growing worldwide effort towards miniature electronic implants which can be surgically placed in a minimally invasive manner. Examples of these include deep brain neurostimulators, cochlear and retinal implants [6–8]. However, one remaining challenge which has hindered efforts towards miniaturization is centred on methods of delivering power to these implants. Whilst surgical techniques and the ASIC technologies are well evolved to address the requirements of minimally invasive implants, methods of delivering energy to the implant have broadly not followed this trend. Currently, the majority of implants dedicate a significant portion of their volume to power harvesting or storage units [9]. In this work, we report on optical power delivery to a retinal stimulator entirely housed within a diamond capsule. Moreover, the light intensity modulation is used as a means of data telemetry to the stimulator. This approach enables an implant with small dimensions to be wirelessly operated.

Whilst low power prostheses, such as pacemakers with sub-100s µW consumption, can use implanted non-rechargeable batteries to meet their energy needs [10], higher power consumption prostheses (e.g. cochlear or visual implants) require power delivery from external sources. The two widely used methods for continuous power delivery to an implant are (i) percutaneous plugs [11–14] and (ii) inductively coupled resonance coils [15–17]. Percutaneous plugs are generally deemed undesirable due to infection risks associated with the break in the skin's protective barrier [13,14]. Wireless power delivery eliminates this infection risk. Wireless systems using inductively coupled resonance coils are the prevalent approach in marketed prostheses such as Argus II retinal stimulator [18]. Despite their high efficiencies, geometric constraints of these coils mean that they characteristically occupy relatively large volumes to provide sufficient power to the implant. Various approaches have attempted to address these challenges. For example, ultrasonically powered and controlled stimulators such as StimDust have been developed, with a volume of 6.5 mm$^3$ capable of receiving 145 µW of electrical power at a depth of 21.5 mm below the skin [19]. Alternatively, power can be delivered to an implant with a laser and converted to electrical power using photovoltaic (PV) cells leading to increased power densities through volume reduction. This approach is useful if the implant is close to the tissue surface, or if there is an optical path from the laser to the implant such as in the eye. Indeed in our earlier work, we have demonstrated a power density of ~20 mW/mm$^3$ using optical approaches, which is about one order of magnitude larger than what is achievable using inductively coupled resonance systems [20]. The PRIMA sub-retinal device [21] is an example of the optically powered and controlled retinal stimulator which is undergoing clinical trials. Each pixel of the PRIMA device acts as a photocell directly converting light pulses into electrical signals which stimulate bipolar cells. The device described here differs from the PRIMA device in that laser beam is used as a power and digital data source for an ASIC embedded in the stimulator package. This provides maximum possible flexibility in terms of stimulation strategies with the pulse duration, current, polarity, and timings all digitally controlled.

In the case of a retinal implant, the stimulator needs to be packaged as a miniaturised implant to fit within the confines of the eye in a minimally invasive manner [22,23]. If coils are to be used for power delivery, they typically will exceed the size of the eyeball [24], requiring the placement of the coils outside of the eyeball, at proximity to stimulating electrodes positioned inside the eye. In this case, the electrical wires are required from the coil to the device. These wires which breach the eyewall, and need to remain in place permanently to operate the implant. As well as an increase in the complexity of the surgical procedure used to place the implant in the eye, this results in an elevated risk of hypotony or conjunctival erosion [25]. It should be noted that devices such as Argus II successfully use this approach, and through sophisticated engineering and surgical methods have substantially reduced these risks [25,26]. Nevertheless, there have also been attempts to eliminate the wire crossing the eyewall by placing the receiver coil inside the eye. For example, the 25 electrode RET-3 retinal stimulator places the receiver coil and electronics within an artificial lens in the posterior chamber of the eye [7]. However, a comparison of this method with other schemes is difficult because the power consumption of the implant has not been fully reported.

In this work, we employ optical power and data transmission for a retinal implant designed to target degenerative vision conditions such as retinitis pigmentosa. Although patients with these types of conditions have lost their photoreceptive cells, the remainder of the nervous system responsible for conveying visual information to the brain remains intact. In a healthy retina, the photoreceptors convert optical input to an electrical charge. These are detected by bipolar cells which in turn transmit the visual information to the retinal ganglion cells. Retinal stimulation implants aim to replace the role of degenerated photoreceptors by electrically stimulating the retinal nerve endings [27]. Devices such as Argus II utilise this mechanism to stimulate the ganglion cells in an epiretinal configuration using an array of 6 x 10 electrodes. The quality of the vision provided by the retinal implant is determined, to a degree, by the number of stimulating electrodes [23,28,29]. There are approximately 16,000 retinal ganglion cells per $mm^2$ at an eccentricity of 2 mm from the fovea [30], each providing a potential input site for stimulation. Fabrication of arrays of electrodes of a density corresponding to the number of ganglion cells per $mm^2$ in the macula is extremely challenging, partially due to the spread of current from an individual electrode. An alternative to simply increasing the density of electrodes is to improve the selectivity of the stimulation, using techniques such as current steering and current focusing [31,32]. Implementation of advanced stimulation strategies requires the development of state-of-the-art ASIC devices which allow flexibility in the stimulation strategy [33]. The stimulator also needs to be packaged as a miniaturised implant to fit within the confines of the eyes in a minimally invasive manner [22,23]. Thus, there is a strong motivation for a small implant with no wires, but one that contains an ASIC capable of implementing smart stimulation strategies. One obstacle to achieving this goal has centred on the size limitation imposed by traditional power delivery schemes. As mentioned above, and, notwithstanding some attempts to place coil inside the eye, the most common approach is inductively coupled wireless power transmission, which utilises coils with dimensions that exceed the size of the eyeball [24] and therefore, require a wire that breaches the eyewall.

An alternative approach is based on the observation that the eye provides the ideal location for an optically powered implant given its small size, delicate structure, and direct optical pathway from the retina to the outside world. As mentioned above, existing optical approaches have focused on the use of photodiode arrays to directly generate electrical stimulation pulses – examples include such as the PIXIUM device and the Retina AG devices [29,34]. In this case, the power/data is delivered to the implant in both spatially and temporally encoded fashion. The desired electrode is activated by directing a small light spot to the target photodiode at a given location. The duration of the light pulse determines the temporal characteristics of the stimulation pulse. However, the absence of digital electronics means that there is limited scope for utilising a wide range of stimulation modalities. For instance, existing optical approaches do not provide for the capability of the on-the-go selection of cathodic leading or anodic leading phase biphasic pulses. They also do not offer precision control of key features of stimulation waveforms such as inter-pulse durations.

Moreover, the task of activating a single, predetermined electrode is challenging. This is because the electrode selection is spatially encoded [35,36] and therefore light needs to be directed to the specific electrode with a precision comparable to the size of that electrode. The challenge arises when considering the inherent eye movements and the subsequent requirement for eye-tracking and compensation mechanism.

In this work, we report on a new generation of retinal implant that is powered and controlled optically as illustrated in Figure 1. We use a stimulator ASIC specially developed for retinal stimulation which provides a high degree of flexibility by allowing the selection of various stimuli waveform parameters and individually addressable electrodes [33]. We combine optical energy delivery with data telemetry to the ASIC. The implant is packaged using diamond technologies specifically developed for the fabrication of high acuity epiretinal prostheses. As well as providing a long-lasting bioinert encapsulation approach, diamond packaging is well suited for high-density 2.5D integration technology, owing to the extensive use of vertical interconnects, resulting in a miniaturised implant. Moreover, the optical transparency of the diamond material enables a direct optical path to the optoelectronic components inside the implant. The stimulator ASIC is connected with a hermetic array of 256 diamond-based microelectrodes [37] using vertical interconnects [23]. In our earlier work, we have reported a diamond based epiretinal stimulator which utilised electrical wires for power and data delivery [22]. The in-vivo and in-vitro attributes of the device [22,38–40], along with the electrochemical performance of the electrodes have been reported [37,40]. The present work builds on our previous reports to demonstrate the feasibility of a wireless miniature implant using optical data and power delivery. The implant benefits from small dimensions, making it possible to place it entirely within the confines of the eye.

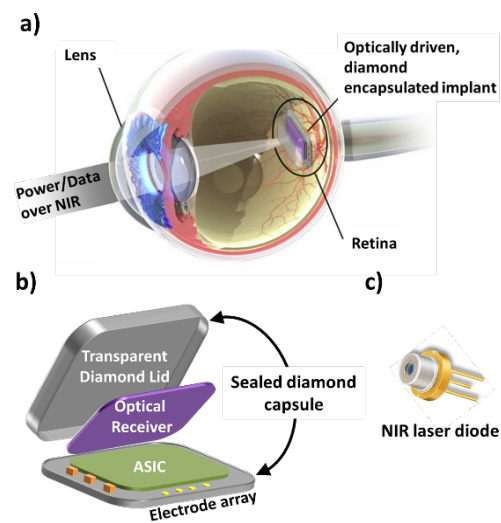

**Figure 1.** a) An illustration of a miniature microelectronic implant which receives optical data and power through the pupil via a near-infrared laser. The implant is designed for epiretinal placement. b) Key components of the stimulator including the simulator ASIC, optoelectronic elements for receiving power and data, diamond electrode array and transparent diamond lid. c) A packaged laser diode used to power the implant (150mW optical power at 850nm).

Section 2.1 provides an overview of the system architecture, including the design of the external hardware and the implant. A detailed description of the implant design is presented in Section 2.2, with the focus on the power harvesting and data extraction approaches. Section 2.3 describes the fabrication of the diamond implant and the packaging of various electronic components inside. The 2.5D integration architecture achieved through the use of a novel diamond interposer technology will be presented. This key feature enables vertical stacking of electronic components resulting in a compact device. The diamond interposer is micromachined for allowing a further reduction in size. Section 2.4 presents the key aspects and design considerations when using PV energy harvesting system. A novel PV cell technology is identified and its characteristics using safe light intensities is reported. The response of the PV cell to intensity-modulated incident light and its potential as an optical data receiver is discussed. Section 2.5 builds on this by presenting an optoelectronic circuit which utilises the PV cell in combination with a photodiode to capture both optical data and power. The fully functional implant consisting of diamond packaged optoelectronic circuit and stimulator ASIC is presented in Section 2.6. The capacity of the implant in delivering stimulation pulses using optical power and data delivery is demonstrated and its performance analysed. The results of this work are discussed in section 3, and the conclusions of this work are presented in Section 4.

## 2. Results and Discussion

## 2.1 System Architecture

This work utilises an existing retinal stimulator ASIC and its associated external driver control circuit. Earlier works have reported on the design of the stimulator ASIC [33], and its use as part of a retinal implant [22]. Some of the key features of the stimulator ASIC are its ability to provide a high degree of flexibility in selecting various stimulation parameters, and the possibility of directly addressing each of the 256 electrodes. The stimulator supports a refresh rate of 60 Hz, and all electrodes can be activated simultaneously. Moreover, the electrodes can be digitally configured as an active electrode or return electrodes. In a retinal implant, these features enable stimulation that can be customised for the patient based on their feedback. In our earlier work, we demonstrated a wired retinal implant consisting of the stimulator ASIC encapsulated in a diamond capsule with an array of diamond-based electrodes [23]. The present work builds on the previous design by demonstrating optical power and data delivery. Because of the high energy density of the PV cell and its compatibility with densely integrated microelectronics, the entire system can be implemented within the confines of the single implant. As shown in the block diagram in Figure 2, the overall system consists of two optically linked units: an external module and an implant module. Power and data are generated by the external module and transmitted to the implant module using intensity-modulated laser light based on an amplitude shift keying (ASK) modulation scheme.

The external module consists of a personal computer (PC), external driver controller, laser driver and a laser diode. The light source selected in this work is a near-infrared laser diode (852 nm). As discussed further in Section 2.4, this illumination wavelength provides the optimum trade-off between the PV cell's peak operation efficiency and the highest illumination intensity which can safely be used. The laser diode is controlled via a laser driver at two predetermined levels which serve to define the intensities used in the ASK modulation. The system is powered using a DC source delivering ~100 mA at 3.7V. The control data is generated by the external driver in the form of a Manchester encoded binary data stream. The specific stimulation commands to the ASIC are generated by the external driver based on the input from the PC.

A PV cell, a photodiode, an interface circuit, and the stimulator comprise the implant module. A high performance vertically integrated multijunction GaAs PV cell converts the incident laser beam to electrical power. Further information on the PV cell characteristics is provided in Section 2.4. A reverse-biased GaAs photodiode is used to generate the data signal based on the ASK-modulated light intensity incident on the implant. A compact interface circuit using commercially available components is developed to power the stimulator ASIC and provide the data signal, which delivers stimulation pulses based on commands initiated by the external PC.

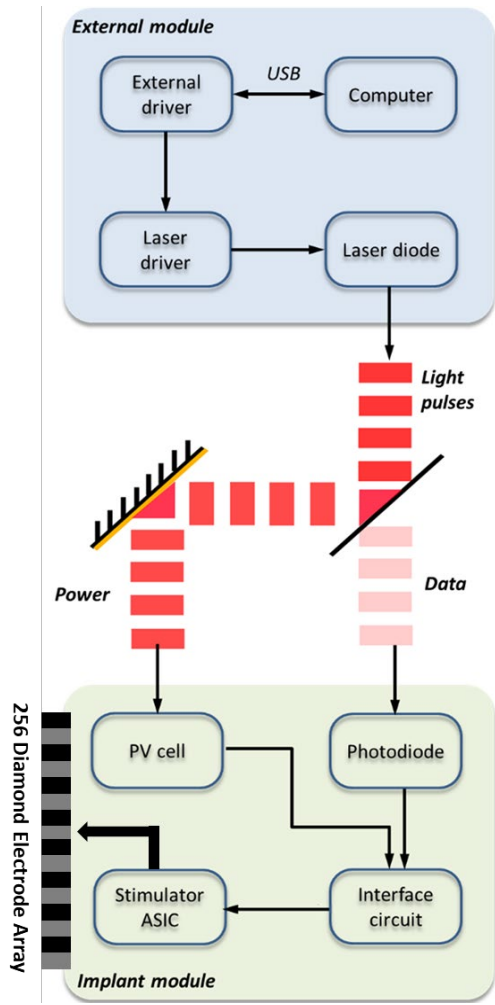

**Figure 2.** Block diagram of key parts of the system implemented in this work. The system consists of an external module and an implant module. The external module supplies the implant module with power and data using an intensity-modulated laser beam. Stimulation commands are conveyed from a computer to an ASIC circuit via a USB connection. The ASIC's digital control unit, hosted within the ASIC stimulator, generates the corresponding data stream, which is optically transmitted to the implant module via a laser diode and driver. The implant module extracts the data and power components and conveys them to the stimulator ASIC via an interface circuit.

## 2.2. Implant unit design

The stimulator ASIC used in this work receives its data and power through a differential pair input signal with a switching frequency of 600 kHz. The input signal is a square wave with rise and fall times of 50 ns and maximum and minimum voltages of 3.6 V and 0 V, respectively. An on-board rectifier is used for AC/DC conversion. A voltage conditioning unit generates ~3 V DC for the analogue circuit elements including the current source drivers and 1 V DC for the digital circuit elements. In parallel to the rectifier and voltage conditioning units, a signal processing unit is used to extract and digitize the 600 kHz Manchester encoded input signal. This is used by the ASIC's data and clock recovery units to decode stimulation commands and generate a 600 kHz clock signal.

Square pulses with 50 ns rise and fall times are used for the operation of the ASIC. Longer rise/fall times results in the generation of a poor quality clock signal and faulty ASIC operation. ASK modulation of the laser light with 50 ns rise/fall time is readily achievable using laser diode driver shown in figure 2. However, the limited bandwidth of the PV cell prevents it from being used to recover a 600 kHz clock signal with 50 ns rise/fall times. To overcome this limitation a dedicated photodiode was used. In this configuration, the PV cell assumes the role of energy generation while the photodiode provides signal demodulation. The photodiode was configured in the reverse biased condition using the voltage generated by the PV cell, and connected in series with a sense resistor. The ASK-modulated optical data was captured as changes in the photocurrent and detected as voltage pulses across the sense resistor. These data pulses were used to operate the ASIC via an interface circuit consisting of three Schmitt trigger inverter logic gates. The first gate was configured as an amplifier, with the subsequent two providing the differential signal to the ASIC. Further information on the design of the interface circuit is provided in Section 2.5. The use of the inverter as the interface circuit provides a significant space-saving benefit whilst using commercially available circuit components. Indeed, bare die hex inverters are readily available and provide the required logic gates in a 1x1 mm sized die. The interface circuit could then be fabricated and packaged along with the rest of the circuit elements within a miniature implant without the need to redevelop the stimulator ASIC. Figure 3 shows the active components used in the fabrication of the implant module.

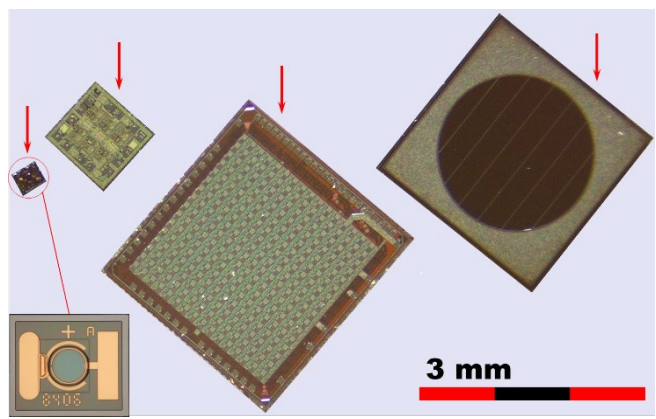

Figure 3. Microphotograph of the implant module's active components. The components from left to right are: photodiode, Schmitt trigger logic IC, retinal stimulator ASIC with 256 active electrodes, and the PV cell. Bottom left insert is the magnified image of the photodiode, highlighting the circular photoactive centre. The scale bar is 3 mm.

## 2.3. Implant fabrication

The implant module consists of two vertically stacked sub-assemblies as depicted in Figure 4 a). Both the interposer and electrode array subassemblies were populated on commercially purchased insulating polycrystalline diamond (PCD). The electrode array subassembly was constructed using a PCD plate with a thickness of 200 µm. The electrode array subassembly hosted the diamond based electrodes, in the form of conductive nitrogen doped ultrananocrystalline diamond (N-UNCD) electrodes on the outer side (bottom). Some of the electronic devices, including the ASIC was placed in the inner side (top) of the electrode array subassembly. The N-UNCD was deposited using microwave plasma chemical vapour deposition (MPCVD) using method outlined in earlier works [36]. N-UNCD is a mixed phase material consisting of nanometre sized crystals (2~5 nm) and graphitic grain boundary. It exhibits a favourable room temperature conductivity [41,42] and excellent electrochemical properties for neuromodulation and recording [40,43]. The thickness of the N-UNCD film was 20 µm and the deposition time 36 hours. The N-UNCD film was fashioned into 256 electrodes using laser micromachining. The electrodes were 120 µm × 120 µm in size and pitched at 150 µm. The electrochemical impedance at 1KHz averaged across all 256 electrodes was 3.1 KΩ (see section 2 of the supplementary information for further detail). In earlier works, the N-UNCD electrodes have been used to deliver electrical stimuli to the retina [38]. The connection between the N-UNCD electrodes and the stimulator ASIC was made by the mean of vertical feedthroughs across the thickness of PCD, and indium-based flip-chip bonding process[23]. More information on the diamond electrode array and connection to the stimulator ASIC is provided in section 2 of the supplementary information.

The second subassembly was a diamond interposer which was populated with the PV cell, photodiode and additional passive components using the fine placement tool. The interposer was made of insulating PCD with a thickness of 500 μm – TM100 E6. The metallisation approach used for the diamond electrode array sub-assembly was used to incorporate metallic tracks within the interposer. The reverse side of the interposer subassembly was milled using laser micromachining to create cavities which accommodated the topology of the electrode array subassembly. This approach facilitated a reduction in the thickness of the implant by placing the stimulator ASIC, along with additional components within the interposer subassembly's cavities. The electrical connection between the interposer subassembly and electrode array subassembly was achieved using vertical feedthroughs. Further detail on the device assembly steps is provided in section 3 of the supplementary information.

The fully assembled diamond device is illustrated in Figure 4b. The device shown here does not incorporate the diamond lid and is connected using wire bonds to a PCB for bench testing. The lateral implant dimension achieved here is 4.6 mm × 3.7 mm with a thickness of 0.9 mm. The 2.5D integration approach achieved through the use of vertically stacked subassemblies enabled a significant reduction in the lateral area occupied by the implant. The thickness of the implant is dominated by the thickness of the active electronic components as well as the two diamond substrates. Although the stimulator ASIC and PV cell each have an approximate thickness of 300 μm, the active device thicknesses in both cases are limited to the top portion of the device. Indeed existing works have already demonstrated that the thickness of these devices can be reduced without significant degradation in their performance [44]. This suggests that there is a path for further reduction in the implant thickness.

In addition to the diamond implant, an identical implant circuit was fabricated on a ceramic chip carrier as shown in Figure 4c. This approach facilitated direct access to the stimulator ASIC's contact pads, a subset of which were connected to the chip carrier. The chip carrier was mounted on a PCB which hosted the passive circuit elements. This test structure provided direct electrical access to the stimulating electrodes for detailed characterisation of the circuit performance.

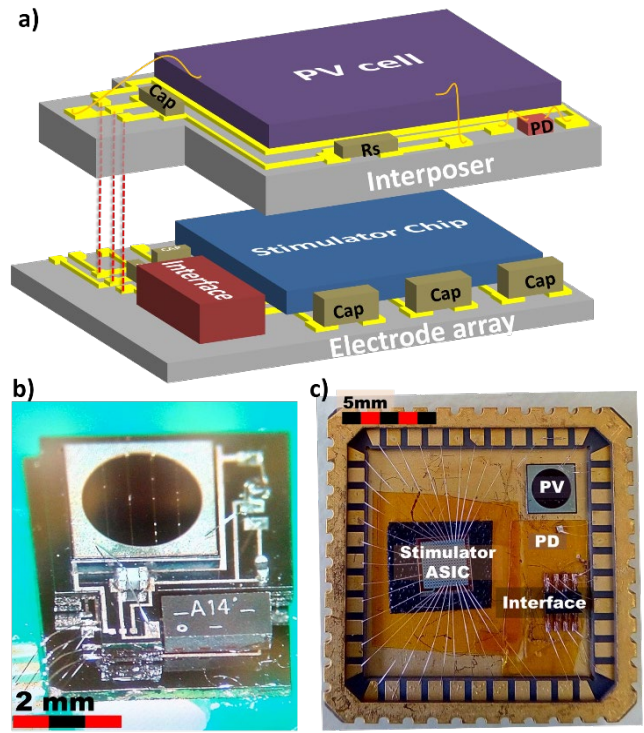

**Figure 4.** a) Schematic illustration of the implant module packaged using diamond technology. The module consists of two substrates; diamond electrode array (lower module) and diamond interposer (upper module). The diamond electrode array hosts the stimulator ASIC, and logic integrated circuit (interface). The interposer hosts the PV cell and photodiode (PD). Additional passive components are used in both substrates. The two substrates are connected using vertical vias. b) Microphotograph of the implant module fabricated using diamond technology and connected to a test PCB board for characterisation. c) The implant module fabricated on a conventional ceramic chip carrier. This configuration allows direct access to the simulator ASIC contact pad electrodes, facilitating direct measurement of the ASIC's output.

## 2.4. PV cell characterisation

The PV cell used in this work fulfils a number of requirements. Firstly, the vertically integrated stack of the PV junctions provides sufficiently large voltages for the operation of the circuit in a space efficient manner. In this work, a PV cell with a stack of five junctions was used, providing a nominal voltage of 5 V. The SEM image shown in section 4 of the supplementary information section highlights the multilayer nature of the PV cell. Secondly, the PV cell selected for this work exhibits an ultrahigh power conversion efficiency of 60% at the wavelength of ~ 850 nm [45], significantly higher than many other state-of-the-art solar cells. By operating such a PV cell with monochromatic light instead of a broadband spectrum, the two fundamental loss mechanisms (thermalisation losses and photon losses) encountered by conventional solar cells are largely eliminated. These factors lead to high external quantum efficiency (EQE) at the wavelength of interest and subsequently higher power conversion efficiency. Thirdly, the PV cell used in this work is optimised for operation at near-infrared (NIR) wavelengths with the peak EQE at 830~850 nm. Operating at NIR wavelengths allows for higher light intensities to be used safely within the eye. Overall these benefits result in a safe high-efficiency power delivery system, which is an essential prerequisite for the use of PV cells in a miniature implant.

In this work, we have used a laser light source with a centre wavelength of 852 nm to deliver power and data to the implant. This wavelength was identified as the optimum operating wavelength by considering three different design constraints: a) the maximum permissible light intensity into the eye, b) optical transmission through the pupil, and c) the EQE of the PV cell. All three design constraints are wavelength dependent and their multiplication product indicates the system's power delivery capacity at a given wavelength. This multiplication product is defined as the power delivery capacity and is used to identify the optimum operating wavelength.

The maximum permissible safe light intensity illuminating an area of 9 mm$^2$ was calculated as a function of wavelength and illustrated in Figure 5a. (See the supplementary information section for further detail). With longer wavelengths in the range of from 600 nm to 950 nm, the maximum safe illumination intensity increases. The optical transmission through the various tissues and fluids of the eye is wavelength dependent. In the case of an adult human eye, the light intensity travelling across the eye and arriving at the anterior surface of the retina is shown in Figure 5b [46]. As highlighted, optical transmission across the wavelength range is near-constant until 850 nm where the transmission decreases rapidly. The main source of this reduction is the optical absorption in the eye's aqueous humour which is predominantly water with a characteristic absorption peak at 980 nm.

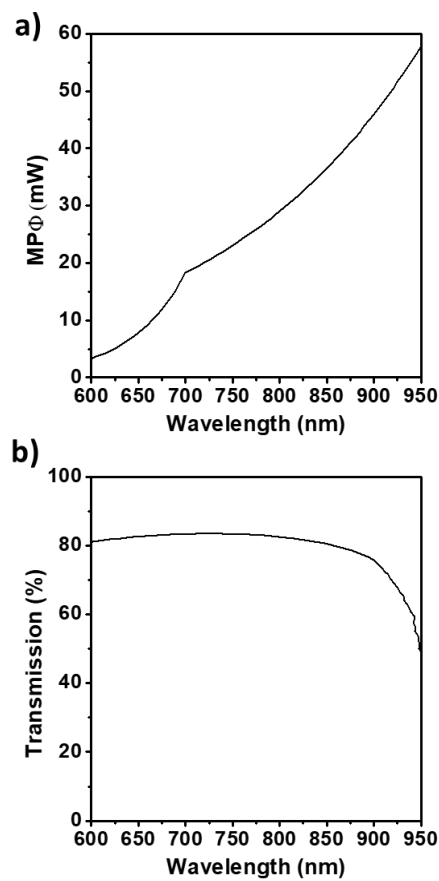

**Figure 5. a)** Dependence of the maximum permissible intensity, $MP\Phi$, on wavelength calculated based on a 3×3 mm light spot size arriving at the surface of the retina. **b)** Optical transmission through the eye to the anterior surface of the retina.

The EQE of the PV cell is depicted in Figure 6 a). This PV technology specific parameter denotes the portion of illumination (photon numbers) which is converted to electrical charge (electron numbers) at a given wavelength. The device exhibits peak responsivity at ~850 nm. There is a sharp reduction in the EQE at wavelengths longer than 850 nm due to the bandgap of GaAs. The EQE exhibits a more gradual reduction between 700 nm and 850 nm. This can be attributed to a photocurrent mismatch between junctions in the PV cell. The thicknesses of the photoabsorber layers in each of the five segments are designed such that at ~850 nm, an equal amount of light is absorbed in each segment and consequently equal photocurrents are generated. At shorter wavelengths, a large proportion of the incident light is absorbed in the upper segments. However, only a small proportion of light reaches the lower segments. As the segments are serially connected, the photocurrent generated by the PV cell is limited by the photocurrent generated by the bottom segment. This results in an EQE reduction at shorter wavelengths. The abrupt nature of this reduction is due to the power-law dependence of photon absorption above the bandgap of the semiconductor.

The optimum operational wavelength of the system can be identified by considering the power delivery capacity of the system as a function of the wavelength as shown in Figure 6 b). This parameter is the numerical product of EQE, $MP\Phi$ and transmission at a given wavelength. As shown, the power delivery capacity peaks at 850 nm. This is the optimum operating wavelength for this specific implant accounting for the PV cell technology, transmission across the eye and the relevant safety limits. Based on figure 5 a), at 850 nm the maximum permissible light intensity arriving at the pupil is 36.5 mW. As shown in figure 5 b), only 80% of the incident light is transmitted to the surface of the retina, resulting in maximum light intensity of 29.2 mW being deliverable to the implant. The EQE of the PV cell at 850 nm is 93% (when scaled by the number of junctions). This means that at 850 nm 93% of the incident photons are converted to electrons, which is close to optimum.

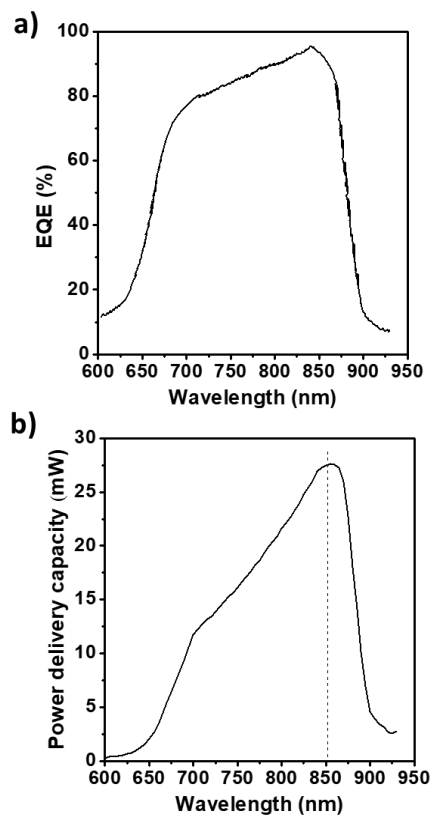

**Figure 6. a)** The EQE (multiplied by the number of junctions) of the PV cell selected for this work. **b)** The maximum safe power delivery capacity of the PV cell as a function of wavelength. Power delivery capacity is defined as the product of $MP\Phi$, optical transmission, and the EQE – is used to identify the optimum operating wavelength of 850 nm.

The current-voltage and power-voltage characteristics of the PV cell under illumination at the intensity of 30 mW are shown in Figure 7a. The voltage at the peak power point is 4.7 V, with an electrical power output of 16.3 mW. This corresponds to an optical-to-electrical power conversion efficiency of 54%. A key feature of high-efficiency PV cells is the high-quality semiconductors used in their fabrication. This translates to long carrier lifetime, which in turn limits their transient responsivity, as depicted in Figure 7b. Here the voltage across the PV cell was measured with a 1.5 kΩ load across the cell. The load selected is the resistive element of the stimulator ASIC's input impedance of 1.5 kΩ. Pulsed laser illumination with a rise/fall times of 50 ns was used. The 70% to 30% voltage rise and fall times extracted from Figure 7b are 370 ns and 190 ns, respectively. The voltage maxima and minima obtained are 3.6 V and 4.25 V respectively. Hence the limited bandwidth provided by the high-efficiency PV cell selected means that a secondary photodiode was required to capture the optically encoded data.

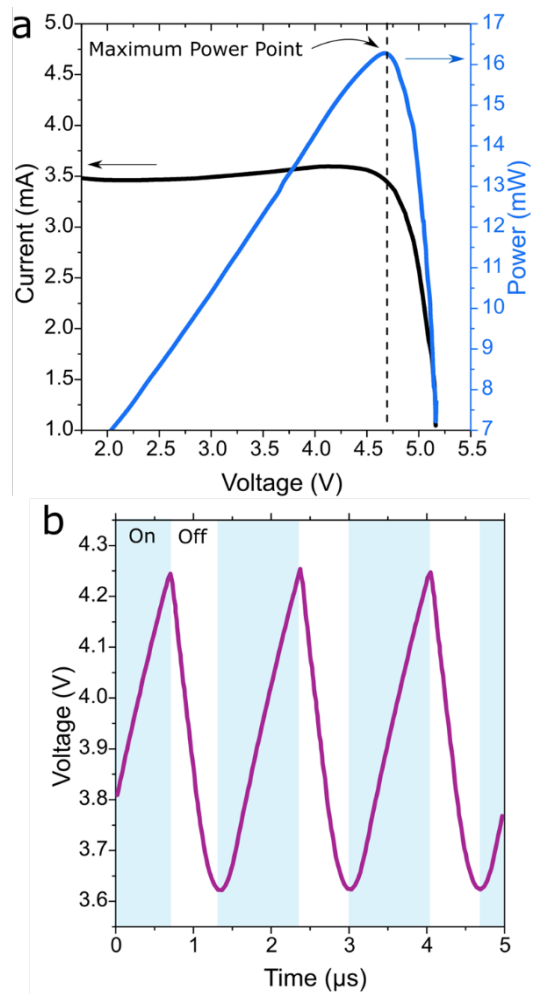

**Figure 7.** a) Illuminated current-voltage and power-voltage characteristics of the PV cell illuminated at an intensity of 30 mW for an 852 nm wavelength. Peak electrical power output of 16.3 mW is extracted at 4.7 V, corresponding to 54% power conversation efficiency. b) Transient voltage characteristics of the PV cell under square pulsed illumination with a load resistance of 1.5 kΩ to emulate the ASIC. The limited rise/fall times shown here are too slow

to enable the extraction of the light intensity-modulated data signal used for the stimulator ASIC operation. This is addressed using a separate photodiode as detailed in Section 2.5.

## 2.5. Optoelectronic interface circuit

Three inverter stages on a hex Schmitt trigger inverter IC were used to amplify and digitise the data signal detected with the photodiode. Unlike the PV cell, these are silicon devices, fabricated using VLSI technology and can be integrated with the ASIC. Figure 8a shows the schematic of the circuit used. Light is directed to both the reverse biased photodiode and the PV cell. The PV cell generates the DC supply to operate the IC. The photodiode is serially connected to an 8.2 kΩ sense resistor. Changes in light intensity are detected as voltage pulses across the sense resistor. The voltage across the sense resistor is capacitively coupled to the first inverter stage using a 150 pF capacitor. The first inverter was configured as a class-D amplifier to amplify and digitise the voltage signal across the sense resistor. This was achieved by connecting the output of the inverter to its input via an 820 kΩ feedback resistor. Using an inverter from the advanced CMOS logic family [47], this configuration produces a nonlinear gain amplifier with its output switching between the supply rails. The second inverter buffers the output of the amplifier, with its output providing one of the differential pair inputs of the stimulator ASIC. The third inverter provided the second of the differential pair inputs by inverting the output of the second inverter. The interface circuit is powered using the PV cell, whilst the input signal to the interface circuit is provided by the photodiode. The outputs of the interface circuit, in the form of a differential pair, provides both power and data to the stimulator ASIC.

Using this approach, it was possible to implement a simple amplification and digitisation circuit without the need for any additional active components. The amplifier used here was a class-D amplifier, which simply switches between the maximum and minimum rail voltages without any linear amplification. Moreover, using the inverter IC offers the benefit of readily providing the differential outputs suitable for use with the stimulator ASIC. Figure 8b shows the interface circuit outputs to the stimulator ASIC in response to light pulses incident on the photodiode. Using this approach, rise and fall times of 112 ns and 160 ns were achieved, respectively. There are voltage overshoots at the point of transition between high and low states. One possible reason for this is the reduction in the current consumed at the amplifier stage at the transition point. This results in a reduction in the current drawn from the PV cell and subsequent increase in its voltage output.

The power consumption of the interface circuit was characterised in the open circuit condition (i.e. without any load connected to the output of the circuit). Here an average current of 0.96 mA was supplied to the circuit by the PV cell, at a voltage of 4.98 V. This is equivalent to the power consumption of 4.78 mW. The relatively large power dissipation by the circuit is expected given the amplifier topology selected.

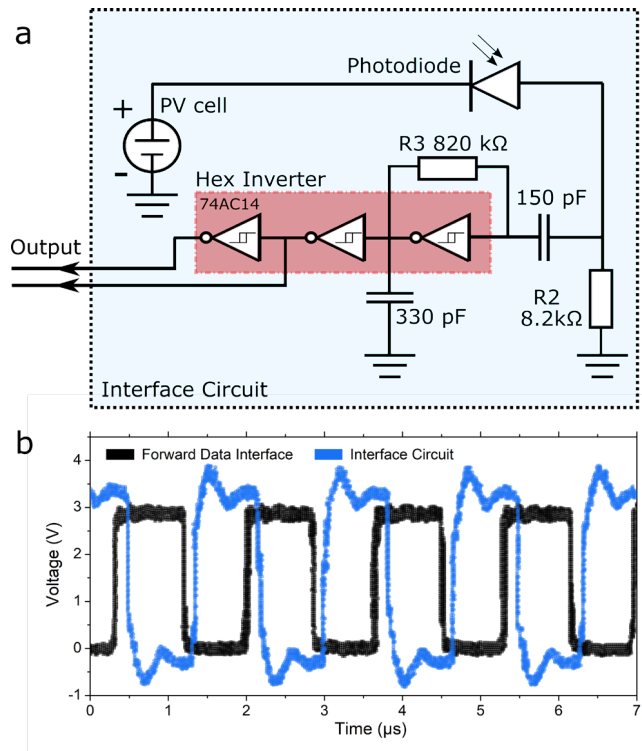

**Figure 8.** a) Schematic of the interface circuit. Series of three inverter gates, implemented using 74AC14 IC, amplify and digitize the signal generated by the photodiode. The IC is powered directly by the PV cell. The interface circuit's output, in the form of a differential pair is used to operate the stimulator ASIC. b) A typical waveform generated by the ASIC driver in the external module compared with the waveform captured and generated by the interface circuit.

## 2.6. Optically driven implant module

The stimulator ASIC was connected to the PV cell and photodiode via the interface circuit described in Section 2.5 to complete the implant module electronics. The implant module was operated using an intensity-modulated laser beam with an average intensity of 24 mW. This is within the safe illumination intensities as discussed in section 2.4 and supplementary information. The laser beam was controlled via a laser driver circuit, with the data to the driver circuit provided by the external driver controller. In order to determine the status of the stimulator ASIC, its back-data channel was monitored using an oscilloscope. The ability of the stimulator ASIC to generate stimulation pulses was determined by connecting one of the electrodes to ground via a 10 KΩ load resistor and monitoring the voltage across the resistor using an oscilloscope.

Upon the start of the illumination, a digitally encoded "power-on-reset" message is generated by the stimulator ASIC on the implant module. This specific data stream, depicted in Figure 9a, is transmitted once the DC voltage at the stimulator ASIC's power recovery unit reaches 3 V. The transmission of the power-on-reset message also requires the ASIC's forward data recovery unit to generate a 600 kHz clock signal. Therefore, the power-on-reset signal confirms that the power and data delivered to the implant module optically is sufficient to both powers the stimulator ASIC and generate a suitable clock signal.

The capability of the implant module in delivering stimulation pulses was determined by programming the stimulator ASIC to generate a stimulation pulse on the target electrode. The stimulation command was programmed on a PC and transmitted optically from the external module to the implant module. Figure 9b illustrates a typical voltage generated across a load resistor for a stimulation pulse. In this case, a biphasic constant current pulse with a duration of 500 µs and amplitude of 250 µA was selected. The voltage measured across the 10 kΩ load resistor was 2.7 V, 0.2 V higher than 2.5 V voltage expected under this condition. This can be attributed to a mismatch of bias currents for the digital to analogue converter (DAC) from the global bandgap reference. As shown earlier this had led to a DAC output current slightly higher than intended. It should be noted that the precise shape of Figure 9 b) will deviate in the physiological environment which includes a capacitive element.

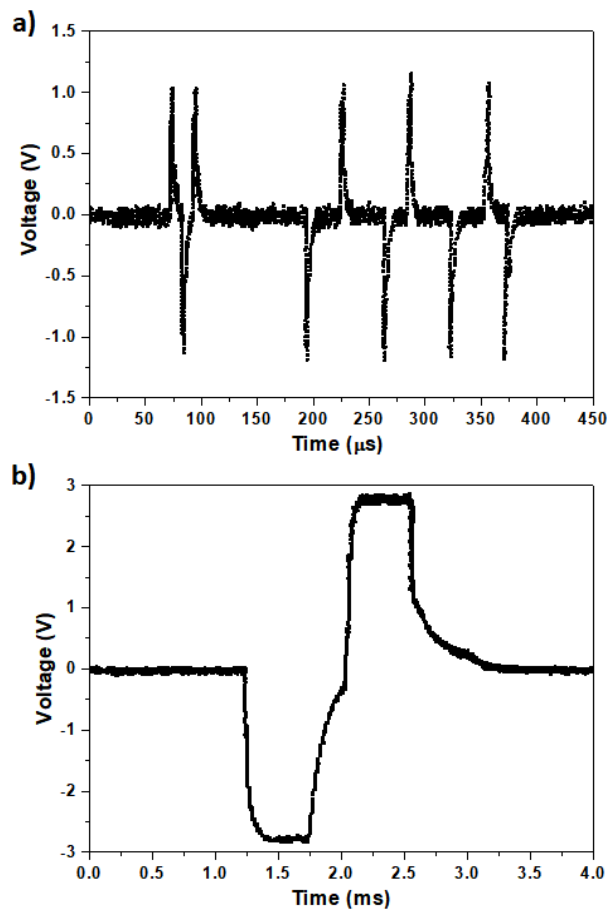

**Figure 9** a) Power-On-Reset message generated by the stimulator ASIC confirming correct supply of power and data signal to the ASIC. b) A biphasic stimulation pulse generated by the stimulator ASIC in response to an optically transmitted stimulation command.

The base power consumption of the implant unit was characterised by measuring the voltage and current generated by the PV cell, with an output load consisting of the stimulator ASIC via the interface circuit. The stimulator ASIC was configured to disable all its stimulating electrodes to assess the base power. In this condition the current generated by the PV cell is 2.98 mA at an operating voltage of 3.4 V, corresponding to electrical power of 10.1 mW. The base power consumption of the stimulator ASIC operated using lead wires had been previously reported as 10 mW [33]. This suggests an increase of 0.1mW in the base power consumption, confirming that the power consumption of the optically driven implant unit is similar to that of the stimulator ASIC alone, with negligible power dissipated by the interface circuit. The average light intensity used to operate the implant unit was 24 mW, leading to 42% optical-to-electrical power conversion efficiency of the PV cell. The PV cell is operated far from its peak power operating point of 4.7 V, which results in a reduction in its power conversion efficiency. In future works, this can be addressed by using a PV cell with 4 junctions. This strategy will provide a lower peak-power voltage that is better matched to the system's operating voltage.

## 3. Discussions

The device described in this work has the dimensions of 4.6 mm × 3.7 mm × 0.9 mm. Beyond this, there is a need for additional support structures which are used during the surgical placement. A silicone housing, used to anchor the device in the epiretinal space, constitutes the largest part of the implant. In our earlier work we reported a wired version of the diamond retinal implant with similar dimensions [22]. Further information on the silicone housing as well as the surgical placement has been provided in our earlier work [22,39]. Beyond some minor changes to the form factor, the key difference between these earlier reports and this work is that the silicone housing no longer contains the electrical wires. These wires represented a significant surgical challenge because they needed to breach the eyewall, and in this work we successfully eliminated the need for them. Whilst the addition of the silicone housing around the diamond device results in increased implant dimensions, the design constraints of the silicon housing is sufficiently flexible to enable simplification of the surgical approach.

The device presented in this work is specifically designed as a high acuity retinal implant. The target placement is in the epiretinal space, over the fovea centralis. This part of the retina is responsible for the high acuity vision and boosts the highest density of retinal ganglion cells. The area is approximately 2.5~3mm diameter which matches the 1.6 x 1.6 mm area covered by the diamond electrode array. Similar high acuity retinal implants using a large number of closely spaced stimulators have been demonstrated. Examples of these include Alpha IMS and AMS, PRIMA, and IRIS II [48–51]. These all use a large number of closely spaced electrodes. In the case of IMS and AMS and PRIMA, a digital control unit is absent, which results in some limitation on the possible stimulation strategies[21,50]. PRIMA is a subretinal device and utilises NIR illumination for both power and data, and therefore does not include a wire crossing the eyewall, whereas both Alpha IMS and AMS are optically controlled but require wires crossing the eyewall to deliver electrical power. Similarly, whilst IRIS II does utilise a digital stimulator, it also requires electrical wires crossing the eyewall. In this work we demonstrate the feasibility of an optically powered/controlled device, with that of a digital control unit on an ASIC with no wires. Alternative strategies for retinal stimulation have been developed which do not target high acuity vision. These types of devices typically use a limited number of larger electrodes which cover a large area of the eye. These devices focus on providing a wide view vision. Most notable examples of this technology are the Argus II retinal stimulator [26] and more recently the BVT device [52].

The use of diamond technology for encapsulation offers several benefits. In our earlier works, we have demonstrated that a hermetic seal can be formed between a monolithic array of diamond electrodes and the implant's body. These electrodes can be as small as 15 μm [40]. The nature of the diamond at the electrode/body interface is such that this seal is robust, inert and biocompatible [37]. Building on this, we also demonstrated the use of active brazing alloy for diamond lid attachment to form a complete capsule [53]. As demonstrated in our earlier work transparent diamond can be used as a lid thus enabling an optical window into the implant [20]. There are different grades of polycrystalline diamond, and their optical performance ranges from the opaque to fully transparent [54]. As presented in the supplementary information section 5, the optical transmission across the diamond lid is 63% when measured in air/diamond/air configuration at the normal angle of incidence. The calculated value of transmission across a single crystal diamond plate in this configuration is 69%. Changing the arrangement to aqueous humour/diamond/air, the transmission further increases to 76%. In the case of the transparent diamond, the main optical loss mechanism is due to the reflectivity at the water/diamond interface. In the case of aqueous humour/diamond the reflection is approximately 8%. The air/diamond interface exhibits a larger reflexion of 17%. The transmission is a function of incidence angle. As shown in the supplementary information section 5, the angle of incidence on the implant is between 0 to 5 degrees. In this range, the variation in the calculated or measured transmission is below 1%.

There are a number of method which can be utilised to increase the transmission across the diamond window. These are centred around minimising the reflection at the two interfaces through the minimising the refractive index mismatches. Two key methods are anti-reflective thin films coatings, consisting of a CVD deposited films such as Silicon Oxide or Nitride, and graded refractive index using nanotexturing. The thin-film coating is suitable for the internal surface of the implant, whilst nanotexturing is useful for the external surfaces. In future works, we will report on a method of reducing this further. And because the absorption of the diamond itself is minimal, the transmission is essentially independent of the lid thickness.

The AISC device used here offers several key features which have been discussed in detail in earlier works [33]. Briefly, these can be categorised as functional or safety features. Among the functional features is the ability to configure any of the 256 electrodes to operate as either an active electrode or return electrode. Moreover, the stimulation waveform can be configured as biphasic or monophasic with cathodic or anodic as the leading phase. The timing features of the stimulation pulse such as the duration of each phase and interphase duration are configurable. Among the safety features are temperature monitoring, compliance voltage monitoring and active charge balancing. Moreover, the AISC provides feedback data on its operation states.

In this work, we have used an optical power level which is within the safety limits set by maximum permissible exposures for ocular safety (ANSI 2000) [55]. The calculations are outlined in the supplementary information section and are based on continuous exposure of 8.3 hours over a 9 mm$^2$ area of the retina. Beyond this safety standard, epidemiologic studies have investigated the impact of NIR and IR eye exposure for industrial workers, prior to the adoption of personal protective equipment as standard working practice. For example, it has been reported that some glass and steelworkers develop cataracts as the result of daily exposure to IR light at intensities in the range of 80 to 400 mW/cm$^2$ over 10 to 15 years [56]. In this work, based on the safety standard we have determined the maximum exposure limit of 36.5 mW entering the pupil. Normalised to a 5mm diameter pupil, this corresponds to 171 mW/cm$^2$. This is comparable or lower than the intensity which has been experienced by the industrial workers.

In addition to the maximum permissible illumination intensity, the implant's electrical power dissipation is an important safety factor. The main damage mechanism from the NIR illumination, at the relevant intensities and durations, is a thermal one. Similarly, the tissue damage as the result of excessive electrical power dissipation is a thermal process. Nevertheless, there is an important difference between the two. NIR's long penetration length means that the light absorption takes place across the depth of the tissue, whilst electrical power dissipation is concentrated at the surface of the device. This means NIR exposure results in a temperature rise across the depth of the tissue, whereas electrical power dissipation results in temperature peaks at the retina-device interface. In addition to the optical safety limits, the electrical power dissipation of the implant is a critical design constraint. An earlier study has demonstrated a maximum power dissipation of 1.46 mW/mm$^2$ can be accommodated to avoid 2 °C temperature rise [57]. The study was performed using an epiretinal implant with dimensions of 5 × 5 mm placed in direct contact with the retina. The device consisted of copper coils on a 159 μm thick polyimide PCB. Applying this limit to the 4.6 × 3.7 mm diamond device, a maximum electrical power dissipation limit of 24.8 mW can be deduced.

However, this is an underestimate because the high thermal conductivity of the diamond (1kW /K.m – TM100, manufacturer datasheet) compared with the polyimide means the heat dissipation is distributed through the 3D implant's exposed surface. The low thermal conductivity of the PCB supports the assertion that the heat is largely dissipated through the surface of the device in contact with the retina, as opposed to the backside of the PCB. Because of this, the polyamide device can be treated as a 2D object. However, in the case of a diamond implant, the heat is uniformly dissipated across a larger area through the front, back and the sides of the implant. Following this, an upper limit on power dissipation can be calculated, using the total area of the diamond implant, 48.98 mm$^2$, leading to an upper limit of 71.5 mW.

**Figure 10** illustrates various mechanisms for optical power loss and electrical dissipation in the device. The electrical power is dissipated by the ASIC, the interface circuit as well as the PV cell. In our work we have measured a power dissipation of 10.1 mW by the ASIC and associated circuits. However, this does not include the dissipated power by the PV cell in the process of converting the incident light into electrical energy. In this work the diamond device was operated using a NIR intensity of 24 mW. If there were no optical losses, this means that a total of 24 mW is dissipated by the implant, a portion of which is dissipated by the PV cell and the remaining by the ASIC and interface circuit. However optical losses mean that only a portion of the 24 mW light is absorbed by the PV cell. One possible approach for estimating the optical losses is to consider the EQE of the PV cell – a parameter which indicates what proportion of the incident photos are converted to electrons. Based on this the optical losses can be approximated as ~7% for an EQE of 93% at this wavelength. This suggests that on this only 22.3 mW of the light is internally used by the PV cell, and 1.7 mW is optically lost. The 22.3 mW is dissipated by the PV cell to generate 10.1 mW of electrical energy, which is then dissipated by the ASIC and interface circuit. The 22.3 mW power dissipation is below both upper (71.5 mW) and lower (24.8 mW) estimates for the maximum power dissipation of the implant.

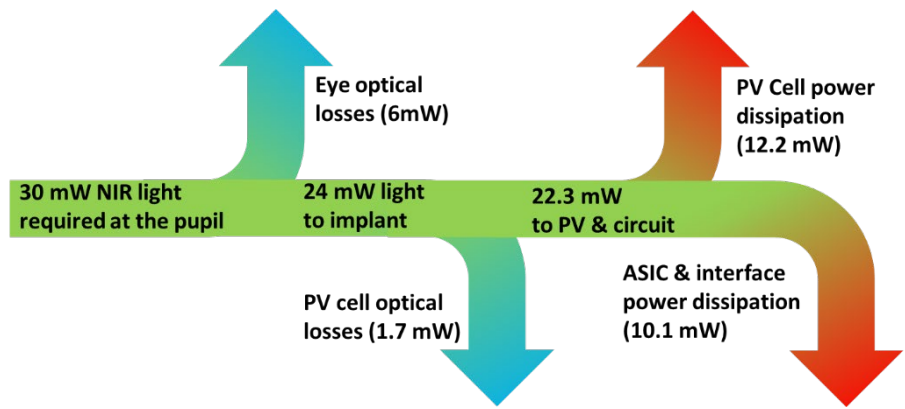

**Figure 10** Implant power budget. Based on 80% transmission from the pupil's surface to the anterior surface of the retina, it is estimated 30 mW of NIR light at the pupil is required to operate the implant. This is below the 36.5 mW optical safety limit. Once inside the eye, we estimate 20% (6mW) of the light is lost, with only 24 mW being available to the implant. Approximately 7% of this light (1.7mW) is optically lost at the PV cell (e.g. reflection) with only 22.3 mW being used by the implant. This is dissipated by the PV cell, interface circuit and the ASIC. Total power dissipation is 22.3 mW.

## 4. Conclusions

This work demonstrates a miniature microelectronic implant which is powered and controlled using a beam of NIR laser light. The feasibility of this approach is illustrated using a diamond packaged retinal stimulator for safe and effective neuromodulation. The large junction capacitance of high performing PV cell used, limits its transient response to light intensity encoded data, which precludes its use for clock recovery and data demodulation. This problem is overcome through the use of photodiodes as the data receiving element and limiting the role of the PV cell to a high performance power transducer. An interface circuit was constructed to provide the stimulator ASIC power and data from the PV cell and photodiode respectively. Using this approach, 10.1 mW of power was delivered to the stimulator ASIC using light intensities of 24 mW at the wavelength of 852 nm. This is within the established optical safety standards. The interface circuit demodulated the optically encoded data and produced signals with rise/fall times of 112/160 nm for the stimulator ASIC operation. The system produces stimulation pulses in line with earlier reports [22]. The implant is packaged using diamond packaging technology for safe and effective neuromodulation, aspects of which has been reported in our earlier works [22].

The resultant implant benefits from small dimensions, making it possible to place it entirely as a thin self-contained unit within the confines of the eye. The use of a laser to power and control the implant enables wireless operation within the established safe light intensity limits in the eye. Compared to existing retinal stimulators which use receiver coils and electronic circuits located outside the eye and require electrode wires to permanently breach the scleral wall of the eyeball, a laser-powered, diamond packaged, miniature implant lends itself to simpler surgical placement procedures, higher efficacy, and superior chronic bio-compatibility. Using an ASIC with a digital control unit, our approach provides a wealth of stimulation modalities and waveforms, required for advanced stimulation strategies to improve vision outcomes. Further work will concentrate on lowering the power consumption of the ASIC and simplifying its input stages to be compatible with the PV and photodiode output thus eliminating the need for the interface circuit.

## 5. Methods

### 5.1 Electronic System

**External system**

The external module consists of a personal computer (PC), external driver controller, laser driver and a laser diode. The light source selected in this work is a 150 mW 852 nm laser diode (M9-852-0150 Thorlabs). The laser diode and its built-in photodiode are connected to a laser driver (iC-NZ, iC-Haus Inc.) to control and monitor the laser light intensity at two predetermined levels which serve to define the intensities used in amplitude-shift keying (ASK) modulation. The laser driver is powered using a benchtop DC source, whilst its control data is generated by the external driver. The specific stimulation commands to the ASIC are generated by the external driver based on the input from the PC running a Matlab code.

**Implant system**

A PV cell, a photodiode, an interface circuit and the stimulator comprise the chip implant module. A 5 junction GaAs PV cell (Azastra Opto Inc. PT5) converts the incident laser beam to electrical power. A reverse biased GaAs photodiode (PD05K1 Albis Opto) is used to generate the data signal based on ASK-modulated light intensity incident on the implant. A compact interface circuit using commercially available components is developed to power the stimulator ASIC and provide the data signal, which delivers stimulation pulses based on commands initiated by the external PC.

### 5.2 Implant fabrication.

The implanted module consists of two vertically stacked sub-assemblies as depicted in Figure 4a. Both the interposer and electrode array sub-assemblies were populated on insulating polycrystalline diamond (PCD) with the thicknesses of 200 µm and 500 µm (TM100 - Element Six (UK) Ltd.), respectively. The 200 µm PCD was used as part of the electrode array, whilst the interposer was fabricated using the 500 µm PCD. The insulating PCDs substrates incorporate ~20 µm thick metallic tracks, ~200 µm and 500 µm thick vertical feedthroughs, and contact pads consisting of a silver brazing alloy (Silver ABA - Morgan Technical Ceramics). The first subassembly was populated by the stimulator ASIC, interface circuit and additional passive components. The reverse side of the subassembly hosted 256 hermetically sealed electrodes using nitrogen doped ultrananocrystalline diamond (N-UNCD) film. The electrode fabrication method has been outlined in earlier works [36]. The electrode dimensions were 120 µm × 120 µm with the pitch of 150 µm. The electrode array subassembly was constructed using 200 µm thick PCD. In earlier works these electrodes have been used to deliver electrical stimuli to the retina. The connection between the N-UCND electrodes and the stimulator ASIC was made by the mean of vertical feedthroughs, and indium-based flip-chip bonding process. Remaining components were placed on the subassembly substrate using a fine placement tool (Fineplacer Lamda, Finetech GmbH) and soldered using $Sn_{42}Bi_{58}$ solder (CR11 Edsyn) by means of reflow soldering.

The second subassembly was a diamond interposer which was populated with the PV cell, photodiode and additional passive components using the fine placement tool. The reverse contact of the PV cell was contacted using conductive adhesive epoxy (CW2400, Chemtronics Inc.), whilst the front contact was connected using the wedge wire bonding method. The photodiode's two contacts were connected to the subassembly substrate using bond wires. The remaining subassembly components were connected using $Sn_{42}Bi_{58}$ reflow soldering. The interposer subassembly was constructed using 500 µm thick PCD. The reverse side of the interposer subassembly was milled using laser micromachining to create cavities which accommodated the topology of the electrode array subassembly. This approach facilitated a reduction in the thickness of the implant by placing the stimulator ASIC, along with additional components within the interposer subassembly's cavities. The electrical connection between the interposer subassembly and electrode array subassembly was achieved using silver alloy filled vertical feedthroughs with the length of 500 µm, and $Sn_{42}Bi_{58}$ reflow soldering.

## 6. Acknowledgement

This research was supported by the Australian Research Council, through Linkage Grant LP160101052, Natural Sciences and Engineering Research Council of Canada (NSERC) grants EGP506146-2016 and RGPIN/05783-2014, and by Ontario Centers of Excellence. Authors gratefully acknowledge insightful discussions with D.J. Garret and K. Ganesan.

## 7. Conflicts of Interest

SP is a shareholder in iBIONICS, a company developing a diamond based retinal prosthesis.

# 8. References

[1]  A. Zhou, S. R. Santacruz, B. C. Johnson, G. Alexandrov, A. Moin, F. L. Burghardt, J. M. Rabaey, J. M. Carmena, R. Muller, *Nat. Biomed. Eng.* **2019**, *3*, 15.
[2]  M. J. Weber, Y. Yoshihara, A. Sawaby, J. Charthad, T. C. Chang, A. Arbabian, *IEEE J. Solid-State Circuits* **2018**, *53*, 1089.
[3]  S. Ha, A. Akinin, J. Park, C. Kim, H. Wang, C. Maier, P. P. Mercier, G. Cauwenberghs, *Proc. IEEE* **2017**, *105*, 11.
[4]  R. Muller, H. Le, W. Li, P. Ledochowitsch, S. Gambini, T. Bjorninen, A. Koralek, J. M. Carmena, M. M. Maharbiz, E. Alon, J. M. Rabaey, *IEEE J. Solid-State Circuits* **2015**, *50*, 344.
[5]  M. M. Ghanbari, J. M. Tsai, A. Nirmalathas, R. Muller, S. Gambini, *IEEE J. Solid-State Circuits* **2017**, *52*, 720.
[6]  O. Majdani, T. S. Rau, S. Baron, H. Eilers, C. Baier, B. Heimann, T. Ortmaier, S. Bartling, T. Lenarz, M. Leinung, *Int. J. Comput. Assist. Radiol. Surg.* **2009**, *4*, 475.
[7]  E. Bloch, Y. Luo, L. da Cruz, *Ther. Adv. Ophthalmol.* **2019**, *11*.
[8]  S.-Y. Chang, I. Kim, M. P. Marsh, D. P. Jang, S.-C. Hwang, J. J. Van Gompel, S. J. Goerss, C. J. Kimble, K. E. Bennet, P. A. Garris, C. D. Blaha, K. H. Lee, *Mayo Clin. Proc.* **2012**, *87*, 760.
[9]  M. J. Kane, P. P. Breen, F. Quondamatteo, G. ÓLaighin, *Med. Eng. Phys.* **2011**, *33*, 7.
[10] A. B. Amar, A. B. Kouki, H. Cao, *Sensors* **2015**, *15*, 28889.
[11] J. K. Niparko, R. A. Altschuler, J. A. Wiler, X. Xue, D. J. Anderson, *Ann. Otol. Rhinol. Laryngol.* **1989**, *98*, 965.
[12] R. P. Michelson, R. A. Schindler, *The Laryngoscope* **1981**, *91*, 38.
[13] B. E. Swartz, J. R. Rich, P. S. Dwan, A. DeSalles, M. H. Kaufman, G. O. Walsh, A. V. Delgado-Escueta, *Surg. Neurol.* **1996**, *46*, 87.
[14] W. Lee, J.-K. Lee, S.-A. Lee, J.-K. Kang, T. Ko, *Surg. Neurol.* **2000**, *54*, 346.
[15] U. Jow, M. Ghovanloo, *IEEE Trans. Biomed. Circuits Syst.* **2007**, *1*, 193.
[16] M. Ghovanloo, K. Najafi, *IEEE Trans. Circuits Syst. Regul. Pap.* **2004**, *51*, 2374.
[17] J. C. Schuder, J. H. Gold, H. E. Stephenson, *IEEE Trans. Biomed. Eng.* **1971**, *BME-18*, 265.
[18] L. da Cruz, J. D. Dorn, M. S. Humayun, G. Dagnelie, J. Handa, P.-O. Barale, J.-A. Sahel, P. E. Stanga, F. Hafezi, A. B. Safran, J. Salzmann, A. Santos, D. Birch, R. Spencer, A. V. Cideciyan, E. de Juan, J. L. Duncan, D. Eliott, A. Fawzi, L. C. Olmos de Koo, A. C. Ho, G. Brown, J. Haller, C. Regillo, L. V. Del Priore, A. Arditi, R. J. Greenberg, *Ophthalmology* **2016**, *123*, 2248.
[19] B. C. Johnson, K. Shen, D. Piech, M. M. Ghanbari, K. Y. Li, R. Neely, J. M. Carmena, M. M. Maharbiz, R. Muller, In *2018 IEEE Custom Integrated Circuits Conference (CICC)*; 2018; pp. 1–4.
[20] A. Ahnood, K. E. Fox, N. V. Apollo, A. Lohrmann, D. J. Garrett, D. A. X. Nayagam, T. Karle, A. Stacey, K. M. Abberton, W. A. Morrison, A. Blakers, S. Prawer, *Biosens. Bioelectron.* **2016**, *77*, 589.
[21] H. Lorach, D. Palanker, In *Artificial Vision: A Clinical Guide*; Gabel, V. P., Ed.; Springer International Publishing: Cham, 2017; pp. 115–124.
[22] A. Ahnood, H. Meffin, D. J. Garrett, K. Fox, K. Ganesan, A. Stacey, N. V. Apollo, Y. T. Wong, S. G. Lichter, W. Kentler, O. Kavehei, U. Greferath, K. A. Vessey, M. R. Ibbotson, E. L. Fletcher, A. N. Burkitt, S. Prawer, *Adv. Biosyst.* **2017**, *1*, 1600003.
[23] A. Ahnood, M. C. Escudie, R. Cicione, C. D. Abeyrathne, K. Ganesan, K. E. Fox, D. J. Garrett, A. Stacey, N. V. Apollo, S. G. Lichter, C. D. L. Thomas, N. Tran, H. Meffin, S. Prawer, *Biomed. Microdevices* **2015**, *17*, 50.
[24] D. D. Zhou, J. D. Dorn, R. J. Greenberg, In *2013 IEEE International Conference on Multimedia and Expo Workshops (ICMEW)*; 2013; pp. 1–6.
[25] L. da Cruz, J. D. Dorn, M. S. Humayun, G. Dagnelie, J. Handa, P.-O. Barale, J.-A. Sahel, P. E. Stanga, F. Hafezi, A. B. Safran, J. Salzmann, A. Santos, D. Birch, R. Spencer, A. V. Cideciyan, E. de Juan, J. L. Duncan, D. Eliott, A. Fawzi, L. C. Olmos de Koo, A. C. Ho, G. Brown, J. Haller, C. Regillo, L. V. Del Priore, A. Arditi, R. J. Greenberg, *Ophthalmology* **2016**, *123*, 2248.
[26] Y. H.-L. Luo, L. da Cruz, *Prog. Retin. Eye Res.* **2016**, *50*, 89.
[27] E. Zrenner, *Science* **2002**, *295*, 1022.
[28] A. Barriga-Rivera, L. Bareket, J. Goding, U. A. Aregueta-Robles, G. J. Suaning, *Front. Neurosci.* **2017**, *11*.

[29] K. Stingl, K. U. Bartz-Schmidt, D. Besch, C. K. Chee, C. L. Cottriall, F. Gekeler, M. Groppe, T. L. Jackson, R. E. MacLaren, A. Koitschev, A. Kusnyerik, J. Neffendorf, J. Nemeth, M. A. N. Naeem, T. Peters, J. D. Ramsden, H. Sachs, A. Simpson, M. S. Singh, B. Wilhelm, D. Wong, E. Zrenner, *Vision Res.* **2015**, *111*, 149.
[30] C. A. Curcio, K. A. Allen, *J. Comp. Neurol.* **1990**, *300*, 5.
[31] C. C. McIntyre, W. M. Grill, *J. Neurophysiol.* **2002**, *88*, 1592.
[32] D. R. Merrill, M. Bikson, J. G. R. Jefferys, *J. Neurosci. Methods* **2005**, *141*, 171.
[33] N. Tran, S. Bai, J. Yang, H. Chun, O. Kavehei, Y. Yang, V. Muktamath, D. Ng, H. Meffin, M. Halpern, E. Skafidas, *IEEE J. Solid-State Circuits* **2014**, *49*, 751.
[34] K. Mathieson, J. Loudin, G. Goetz, P. Huie, L. Wang, T. I. Kamins, L. Galambos, R. Smith, J. S. Harris, A. Sher, D. Palanker, *Nat. Photonics* **2012**, *6*, 391.
[35] D. Palanker, A. Vankov, P. Huie, S. Baccus, *J. Neural Eng.* **2005**, *2*, S105.
[36] A. Asher, W. A. Segal, S. A. Baccus, L. P. Yaroslavsky, D. V. Palanker, *IEEE Trans. Biomed. Eng.* **2007**, *54*, 993.
[37] K. Ganesan, D. J. Garrett, A. Ahnood, M. N. Shivdasani, W. Tong, A. M. Turnley, K. Fox, H. Meffin, S. Prawer, *Biomaterials* **2014**, *35*, 908.
[38] A. E. Hadjinicolaou, R. T. Leung, D. J. Garrett, K. Ganesan, K. Fox, D. A. X. Nayagam, M. N. Shivdasani, H. Meffin, M. R. Ibbotson, S. Prawer, B. J. O'Brien, *Biomaterials* **2012**, *33*, 5812.
[39] K. Fox, H. Meffin, O. Burns, C. J. Abbott, P. J. Allen, N. L. Opie, C. McGowan, J. Yeoh, A. Ahnood, C. D. Luu, R. Cicione, A. L. Saunders, M. McPhedran, L. Cardamone, J. Villalobos, D. J. Garrett, D. A. X. Nayagam, N. V. Apollo, K. Ganesan, M. N. Shivdasani, A. Stacey, M. Escudie, S. Lichter, R. K. Shepherd, S. Prawer, *Artif. Organs* **2016**, *40*, E12.
[40] Y. T. Wong, A. Ahnood, M. I. Maturana, W. Kentler, K. Ganesan, D. B. Grayden, H. Meffin, S. Prawer, M. R. Ibbotson, A. N. Burkitt, *Front. Bioeng. Biotechnol.* **2018**, *6*.
[41] S. Bhattacharyya, O. Auciello, J. Birrell, J. A. Carlisle, L. A. Curtiss, A. N. Goyette, D. M. Gruen, A. R. Krauss, J. Schlueter, A. Sumant, P. Zapol, *Appl. Phys. Lett.* **2001**, *79*, 1441.
[42] J. Birrell, J. A. Carlisle, O. Auciello, D. M. Gruen, J. M. Gibson, *Appl. Phys. Lett.* **2002**, *81*, 2235.
[43] S. A. Skoog, P. R. Miller, R. D. Boehm, A. V. Sumant, R. Polsky, R. J. Narayan, *Diam. Relat. Mater.* **2015**, *54*, 39.
[44] H.-L. Chen, A. Cattoni, R. D. Lépinau, A. W. Walker, O. Höhn, D. Lackner, G. Siefer, M. Faustini, N. Vandamme, J. Goffard, B. Behaghel, C. Dupuis, N. Bardou, F. Dimroth, S. Collin, *Nat. Energy* **2019**, *4*, 761.
[45] C. E. Valdivia, M. M. Wilkins, B. Bouzazi, A. Jaouad, V. Aimez, R. Arès, D. P. Masson, S. Fafard, K. Hinzer, In *Physics, Simulation, and Photonic Engineering of Photovoltaic Devices IV*; International Society for Optics and Photonics, 2015; Vol. 9358, p. 93580E.
[46] E. A. Boettner, J. R. Wolter, *Invest. Ophthalmol. Vis. Sci.* **1962**, *1*, 776.
[47] J. Watson, In *Mastering Electronics*; Watson, J., Ed.; Macmillan Master Series; Macmillan Education UK: London, 1996; pp. 285–293.
[48] K. Stingl, R. Schippert, K. U. Bartz-Schmidt, D. Besch, C. L. Cottriall, T. L. Edwards, F. Gekeler, U. Greppmaier, K. Kiel, A. Koitschev, L. Kühlewein, R. E. MacLaren, J. D. Ramsden, J. Roider, A. Rothermel, H. Sachs, G. S. Schröder, J. Tode, N. Troelenberg, E. Zrenner, *Front. Neurosci.* **2017**, *11*.
[49] R. Hornig, M. Dapper, E. Le Joliff, R. Hill, K. Ishaque, C. Posch, R. Benosman, Y. LeMer, J.-A. Sahel, S. Picaud, In *Artificial Vision: A Clinical Guide*; Gabel, V. P., Ed.; Springer International Publishing: Cham, 2017; pp. 99–113.
[50] K. Stingl, K. U. Bartz-Schmidt, D. Besch, A. Braun, A. Bruckmann, F. Gekeler, U. Greppmaier, S. Hipp, G. Hörtdörfer, C. Kernstock, A. Koitschev, A. Kusnyerik, H. Sachs, A. Schatz, K. T. Stingl, T. Peters, B. Wilhelm, E. Zrenner, *Proc. R. Soc. B Biol. Sci.* **2013**, *280*, 20130077.
[51] M. M. K. Muqit, M. Velikay-Parel, M. Weber, G. Dupeyron, D. Audemard, B. Corcostegui, J. Sahel, Y. L. Mer, *Ophthalmology* **2019**, *126*, 637.
[52] L. N. Ayton, P. J. Blamey, R. H. Guymer, C. D. Luu, D. A. X. Nayagam, N. C. Sinclair, M. N. Shivdasani, J. Yeoh, M. F. McCombe, R. J. Briggs, N. L. Opie, J. Villalobos, P. N. Dimitrov, M. Varsamidis, M. A. Petoe, C. D. McCarthy, J. G. Walker, N. Barnes, A. N. Burkitt, C. E. Williams, R. K. Shepherd, P. J. Allen, for the B. V. A. R. Consortium, *PLOS ONE* **2014**, *9*, e115239.
[53] S. G. Lichter, M. C. Escudié, A. D. Stacey, K. Ganesan, K. Fox, A. Ahnood, N. V. Apollo, D. C. Kua, A. Z. Lee, C. McGowan, A. L. Saunders, O. Burns, D. A. X. Nayagam, R. A. Williams, D. J. Garrett, H. Meffin, S. Prawer, *Biomaterials* **2015**, *53*, 464.

[54] S. E. Coe, R. S. Sussmann, *Diam. Relat. Mater.* **2000**, *9*, 1726.
[55] F. C. Delori, R. H. Webb, D. H. Sliney, *JOSA A* **2007**, *24*, 1250.
[56] E. M. Aly, E. S. Mohamed, *Indian J. Ophthalmol.* **2011**, *59*, 97.
[57] N. L. Opie, A. N. Burkitt, H. Meffin, D. B. Grayden, *IEEE Trans. Biomed. Eng.* **2012**, *59*, 339.